\documentclass[12pt,a4paper]{article}
\pdfoutput=1

\usepackage{amsmath,amssymb}
\usepackage[bookmarks=true]{hyperref}
\usepackage[nosort]{cite}
\usepackage[bulletsep]{collect}
\def\gfxon{\usepackage[final]{graphicx}}

\gfxon

\makeatletter
\let\old@startsection=\@startsection
\renewcommand{\@startsection}[6]{\old@startsection{#1}{#2}{#3}{#4}{#5}{#6\mathversion{bold}}}
\makeatother

\newcommand{\dpou}[1]{\partial^{#1}}
\newcommand{\dpod}[1]{\partial_{#1}}

\makeatletter \@addtoreset{equation}{section} \makeatother

\makeatletter
\let\old@makecaption=\@makecaption
\def\@makecaption{\small\old@makecaption}
\makeatother

\makeatletter
\newcommand{\ellSN}{\mathop{\operator@font sn}\nolimits}
\newcommand{\ellCN}{\mathop{\operator@font cn}\nolimits}
\newcommand{\ellDN}{\mathop{\operator@font dn}\nolimits}
\newcommand{\ellAM}{\mathop{\operator@font am}\nolimits}
\newcommand{\ellK}{\mathop{\smash{\operator@font K}\vphantom{a}}\nolimits}
\newcommand{\ellE}{\mathop{\smash{\operator@font E}\vphantom{a}}\nolimits}
\makeatother

\ifx\genfrac\sdflkaj

\else

\fi
\newcommand{\sfrac}[2]{{\textstyle\frac{#1}{#2}}}
\newcommand{\half}{\sfrac{1}{2}}

\newcommand{\nln}{\nonumber\\}

\newcommand{\earel}[1]{\mathrel{}&\hspace{-2\arraycolsep}#1\hspace{-2\arraycolsep}&\mathrel{}}
\newcommand{\eq}{\earel{=}}
\newcommand{\beq}{\begin{equation}}
\newcommand{\eeq}{\end{equation}}

\def\[{\begin{equation}}
\def\]{\end{equation}}
\def\<{\begin{eqnarray}}
\def\>{\end{eqnarray}}

\makeatletter
\def\mr@ignsp#1 {\ifx\:#1\@empty\else #1\expandafter\mr@ignsp\fi}%
\newcommand{\multiref}[1]{\begingroup
\xdef\mr@no@sparg{\expandafter\mr@ignsp#1 \: }%
\def\mr@comma{}%
\@for\mr@refs:=\mr@no@sparg\do{\mr@comma\def\mr@comma{,}\ref{\mr@refs}}%
\endgroup}
\makeatother

\newcommand{\hypref}[2]{\ifx\href\asklfhas #2\else\href{#1}{#2}\fi}

\newcommand{\secref}[1]{Sec.~\multiref{#1}}

\newcommand{\tabref}[1]{Tab.~\multiref{#1}}

\newcommand{\figref}[1]{Fig.~\multiref{#1}}
\renewcommand{\eqref}[1]{(\multiref{#1})}

\ifx\href\asklfhas\newcommand{\href}[2]{#2}\fi
\newcommand{\arxivno}[1]{\href{http://arxiv.org/abs/#1}{#1}}

\begin{document}

\begin{flushright}\footnotesize
\texttt{ArXiv:\arxivno{0904.0509}}\\
\vspace{0.5cm}
\end{flushright}
\vspace{0.3cm}

\renewcommand{\thefootnote}{\arabic{footnote}}
\setcounter{footnote}{0}
\begin{center}%
{\Large\textbf{\mathversion{bold}
Comments on Holography with Broken Lorentz Invariance}
\par}

\vspace{1cm}%

\textsc{Ivan Gordeli and Peter Koroteev}

\vspace{5mm}

\textit{University of Minnesota, School of Physics and Astronomy\\%
116 Church Street S.E. Minneapolis, MN 55455, USA}

\vspace{7mm}

\thispagestyle{empty}

\texttt{gordeli@physics.umn.edu,\,koroteev@physics.umn.edu}

\par\vspace{1cm}

\vfill

\textbf{Abstract}\vspace{5mm}

\begin{minipage}{12.7cm}
Recently a family of solutions of Einstein equations in backgrounds with broken Lorentz invariance was found \cite{Koroteev:2007yp}.
We show that the gravitational solution recently obtained by Kachru, Liu and Mulligan in \cite{Kachru:2008yh} is a part of the former solution which was derived earlier in the framework of extra dimensional theories. We show how the energy-momentum and Einstein tensors are related and establish a correspondence between parameters which govern Lorentz invariance violation. Then we demonstrate that scaling behavior of two point correlation functions of local operators in scalar field theory is reproduced correctly for two cases with critical values of scaling parameters. Therefore, we complete the dictionary of ``tree-level'' duality for all known solutions of the bulk theory. In the end we speculate on relations between RG flow of a boundary theory and asymptotic behavior of gravitational solutions in the bulk.
\end{minipage}

\vspace*{\fill}

\end{center}

\newpage

\section{Introduction}\label{sec:Intro}

It is an interesting question whether it is possible to construct a reasonable quantum field theory with broken Lorentz invariance.
In high energy physics theories with broken Lorentz invariance appear in different contexts such as ultra high energy cosmic rays \cite{Coleman:1998ti} (one may count this application as a pioneering one in physics with broken Lorentz invariance), braneworld models \cite{Gorbunov:2005dd, Dubovsky:2001fj, Libanov:2005yf, Koroteev:2007yp}, and others. The breaking of Lorentz invariance is manifested in changing of the dispersion relation. For instance, scalar perturbations in model \cite{Koroteev:2009xd} upon proper choice of parameters
obey the following dispersion relations at small momenta
\<
E^2\eq 3p^2+\mathcal{O}(p^3)\,,\quad \text{for the zero mode}\nln
E^2_n \eq a+\frac{b\, n^2}{4\log^2\frac{p}{k}}\,, \quad \text{for higher modes} \,.
\>
Here $E$ is the energy, $p$ is four-momentum, $a$ and $b$ are some constants, $n$ is an integer representing the mode number, and $k$ is the curvature scale.
This model represents an example of a theory with broken Lorentz invariance and is formulated in the background
with the following metric\footnote{Sign convention is changed compared to \cite{Koroteev:2007yp}.}
\[
\label{eq:generalansatz}
ds^2 = -e^{-2\xi kz}dt^2 +e^{-2\zeta kz}d\textbf{x}^2+dz^2\,,
\]
were $t$ is time, $\textbf{x}$ is a three dimensional spatial vector, $z$ is the coordinate along the extra dimension, and $\xi$ and $\zeta$ are parameters which govern the Lorentz invariance violation. The above metric is invariant w.r.t. three dimensional translations and rotations. It can be shown by direct computation of Weyl tensor that the space \eqref{eq:generalansatz} is conformally flat if and only if $\xi=\zeta$. A one-parametric family of static solutions with these symmetries was initially found in \cite{Koroteev:2007yp}. The geometry \eqref{eq:generalansatz} appeared to be a solution of Einstein equations with an anisotropic perfect relativistic fluid as matter. The anisotropy coefficient in this fluid is related to $\xi$ and $\zeta$ in \eqref{eq:generalansatz}.

Another example, the so-called Lifshitz model \cite{PhysRevB.23.4615, PhysRevLett.35.1678}, which is employed in description of strongly correlated
electrons, has the following dispersion relation
\[\label{eq:LifshitzDisp}
E^2=\alpha p^4\,,
\]
where $\alpha$ is a parameter. This model admits the so-called dynamical scaling such that the time
and the coordinates scale in different ways
\[\label{eq:dynamicalScaling}
x\mapsto \lambda^\gamma x\,, \quad t\mapsto \lambda t\,, \quad \gamma\neq 1\,.
\]

In this note we discuss the relationship between recent work by Kachru, Liu and Mulligan \cite{Kachru:2008yh} and two papers \cite{Koroteev:2007yp, Koroteev:2009xd}. The former paper was motivated by theories with Lifshitz-like fixed points which have dispersion relations similar to \eqref{eq:LifshitzDisp}. The authors elaborate on gravity dual description of the $2+1$ dimensional Lifshitz-type model which obeys dynamical scaling \eqref{eq:dynamicalScaling}. Its gravity dual is formulated in $3+1$ dimensions and its background can be represented as a deformation of the four-dimensional anti de-Sitter space.
In \cite{Koroteev:2007yp, Koroteev:2009xd} static braneworld models with broken Lorentz invariance in the bulk are investigated in the background \eqref{eq:generalansatz}. The goal of the present work is to show that the solution constructed in \cite{Kachru:2008yh} can be represented as a part of the one which was earlier obtained and investigated in \cite{Koroteev:2007yp, Koroteev:2009xd}. From this prospective, the paper by Kachru, Liu and Mulligan provides a field theoretical description of the gravitational solution \cite{Koroteev:2007yp}. In this note we show that the two solutions indeed coincide and the papers mentioned above represent the same setup.

The paper is organized as follows. In \secref{sec:Corr} we elaborate on the correspondence between the solutions of the two models. In \secref{sec:GGCorr} we derive bulk propagators for two critical cases corresponding to $\xi=1,\zeta=0$ and $\xi=0,\zeta=1$ in \eqref{eq:generalansatz}. We show that two point correlation functions calculated in scalar field theory obey the proper scaling behavior. Then in \secref{sec:Disc} we make a summary and discuss the infra red behavior of the solutions.

\section{Correspondence of the Solutions}\label{sec:Corr}

In this section we relate the solutions obtained in \cite{Kachru:2008yh} with the ones from \cite{Koroteev:2007yp}. First, we briefly recall the two setups and then make a correspondence.

\paragraph{Solution with perfect fluid.}

Let us briefly recall the main results of \cite{Koroteev:2007yp}. In the context of the KLM model we shall recast the solution of \cite{Koroteev:2007yp} for
the four-dimensional bulk. In order to do that we take the metric \eqref{eq:generalansatz} but with $\textbf{x}$ being a two dimensional spatial vector.
The bulk is filled with perfect anisotropic fluid with the following energy-momentum tensor
\<
\label{eq:EMTperfectfluid}
T^0_0\eq(1+\omega)\rho u_0u^0-p_4\,,\nln
T^1_1=T^2_2\eq(1+w)\rho u_1u^1-p_1\,,\nln
T^4_4\eq(1+\omega)\rho u_4u^4-p_4\,,\nln
T^0_4\eq(1+\omega)\rho u^0u_4\,.
\>
Here $u^A$ is a covariant velocity vector and $u^A u_A=1$. The above formula needs to be completed by the equations of state
\[
p_1=w\rho\,,\quad p_4=\omega\rho\,,
\]
where $p$ is the pressure and $\rho$ is the energy density. The ratio $w/\omega$ plays a role of the anisotropy parameter in this fluid. For $w=\omega$ the above tensor describes isotropic perfect fluid with the equation of state $p=w\rho$. There is also the cosmological constant term $\Lambda$ present in the bulk.
One can show that the bulk Einstein equations\footnote{We use Plank units}
\[\label{eq:Einst}
G^A_B=T^A_B+\Lambda\delta^A_B
\]
are satisfied in the background \eqref{eq:generalansatz} with matter \eqref{eq:EMTperfectfluid} provided that
\<
\label{eq:EinsteinSolutionFluid}
\rho\eq -\Lambda-3k^2 \zeta^2\,,\nln
w\eq -1+k^2\frac{(\xi+2\zeta)(\xi-\zeta)}{\rho}\,,\nln
\omega\eq -1+k^2\frac{2\zeta(\xi-\zeta)}{\rho}\,,
\>
and $u_1=u_4=0$. We see that for the Lorentz invariant case ($\xi=\zeta$) only the bulk cosmological constant remains and  $w=\omega=-1$ corresponding to the equation of state for the vacuum. This is indeed the RS2 model \cite{Randall:1999vf} with shifted cosmological constant $\tilde \Lambda = \Lambda+3k^2 \zeta^2$. Therefore anisotropy of the fluid controls the deviation from the $AdS_4$ configuration.

Note that it is not hard to obtain solutions of the Einstein equations \eqref{eq:Einst} in $d$ dimensions
\<
\label{eq:EinsteinSolutionFluidDdim}
\rho\eq -\Lambda-\half(d-1)(d-2)k^2 \zeta^2\,,\nln
w\eq -1+k^2\frac{(\xi+(d-2)\zeta)(\xi-\zeta)}{\rho}\,,\nln
\omega\eq -1+k^2\frac{(d-2)\zeta(\xi-\zeta)}{\rho}\,.
\>
We can see that in the solution \eqref{eq:EinsteinSolutionFluid} for the null energy condition $\omega>-1$ and $w>-1$ to be satisfied\footnote{See \cite{Koroteev:2007yp} for thorough description of NEC in braneworld scenarios.} one needs to put $\xi>\zeta$.

\paragraph{KLM Solution.}

The KLM solution has the following metric
\[\label{eq:KLMMetric}
ds^2 = L^2\left(-r^{2Z}d\tau^2+r^2d\textbf{X}^2+\frac{dr^2}{r^2}\right)\,,
\]
where $Z$ controls the Lorentz invariance violation and $L$ is the scale. The matter part of the action reads
\[
-\int \frac{1}{e^2} F_{(2)}\wedge\ast F_{(2)}+ F_{(3)}\wedge\ast F_{(3)} - c\int F_{(2)}\wedge B_{(2)}\,,
\]
where $e$ and $c$ are couplings\footnote{The latter is topological and needs to be quantized}. Field strengthes $F_{(2)}=d A_{(1)}$ and
$F_{(3)}=d B_{(2)}$ have the following form
\[\label{eq:KLMMatter}
F_{(2)}=A\,\theta_r\wedge\theta_t\,,\quad F_{(3)}=B\,\theta_r\wedge\theta_x\wedge\theta_y\,,
\]
where $A$ and $B$ are constants, $\theta_r,\theta_t,\theta_x,\theta_y$ are related to $dr,dt,dx,dy$ such that the
metric \eqref{eq:KLMMetric} becomes of the form $\text{diag}(-1\,,1\,,1\,,1)$ in the $\theta$-basis.
The solution of the Einstein equations include
\<
\Lambda \eq -\frac{Z^2+Z+4}{2L^2}\,,\nln
A^2 \eq \frac{2Z(Z-1)}{L^2}\,,\nln
B^2 \eq \frac{4(Z-1)}{L^2}\,.
\>
In order to avoid tachyonic solutions one needs to have $Z>1$.

\paragraph{Correspondence between the solutions.}

First, we observe that the metric \eqref{eq:KLMMetric} coincides with \eqref{eq:generalansatz} up to the following identification rules
\[
t = L\tau\,,\quad \textbf{x}=L\textbf{X}\,,\quad r^2=e^{-2 k z}\,,\quad \xi=Z\,,\quad \zeta = 1\,.
\]
It means the models \cite{Koroteev:2007yp} and \cite{Kachru:2008yh} are formulated in the same background. We can also see that the former is more generic since it allows for the configuration with vanishing $\zeta$\footnote{First the solution $(\xi,\zeta)=(1,0)$ was investigated in \cite{Dubovsky:2001fj}}.

Since l.h.s. of the Einstein equations are the same in both models, we can now explicitly compare the energy-momenta tensors.
Using formulae from the previous paragraph and energy momentum tensor from \cite{Kachru:2008yh} we can summarize the correspondence between the
models in \tabref{tab:Correspondence}
\begin{table}\label{tab:Correspondence}
\begin{center}
\begin{tabular}{|l|c|c|}
\hline			
    &\textbf{KL} & \textbf{KLM} \\ \hline\hline
  Cosmological constant & $\Lambda=-\rho-3k^2$    & $\Lambda=-L^{-2}(Z^2+Z+4)$  \\ \hline
  First component & $w$ & $-1+\displaystyle\frac{A^2+B^2}{2(-\Lambda-3L^{-2})}$ \\ \hline
  Second component & $\omega$ & $-1+\displaystyle\frac{B^2}{2(-\Lambda-3L^{-2})}$ \\ \hline
  LIV parameter  & $w-\omega$  &  $A$                     \\ \hline
  Anisotropy & $p_1-p_4$  &  Energy flux $A^2$ \\
             &             &                        w.r.t $z$ direction  \\ \hline
  Constraints  & Reality of Fluxes  &  Null energy condition  \\ \hline
\end{tabular}
\end{center}
\caption{The form $A_{(1)}$ corresponds to the electric field along the $z$ direction, while the $B_{(2)}$ field corresponds to the isotropic ``magnetic'' field in the $x,y$ and $z$ directions. The presence of the $A_{(1)}$ field provides anisotropy in the bulk and serves as a parameter of the Lorentz invariance violation (LIV). In \cite{Koroteev:2007yp} language the anisotropy is due to different pressures applied along the extra dimension and along the branelike directions.}
\end{table}
%

\section{Two Point Correlation Functions}\label{sec:GGCorr}

\begin{figure}[!ht]
\begin{center}
\includegraphics[height = 10cm, width=11.5cm]{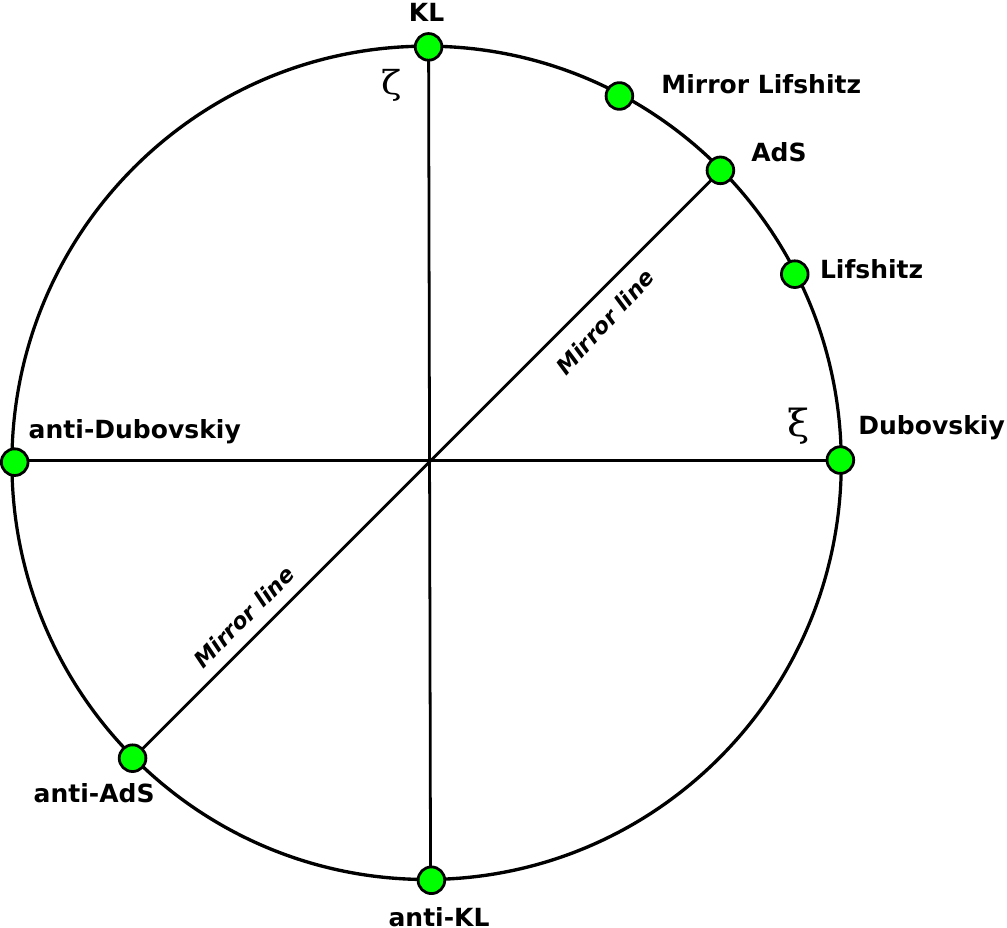}
\caption{The parameter space of metrics. Backgrounds and their mirror duals with known bulk solutions are shown. Although mirror transformation does not make sence in dimensions different from two, on this figure we refer to it as formal interchanging of $\xi$ and $\zeta$ (See \cite{Koroteev:2009qr} for the details). 
Models obtained by reflection of $\xi$ and $\zeta$ are called anti-models. In this note we shall KL and Dubovsky models as critical ones since one of the scaling parameters vanish.}
\label{fig:xizeta}
\end{center}
\end{figure}

In \cite{Kachru:2008yh} a correspondence between the bulk classical action and local operators on the boundary was demonstrated for the Lifshitz theory $(\xi=2,\zeta=1)$ in $2+1$ dimensions. This section generalizes their consideration to other known solutions. In \figref{fig:xizeta} the space of metrics with broken Lorentz invariance along the extra dimension is presented (see also \cite{Koroteev:2009xd}). The known models include 
\begin{itemize}
 \item AdS model $(1,1)$
 \item Lifshitz model $(2,1)$
 \item Dubovsky model $(1,0)$
 \item Mirror Lifshitz model $(1,2)$
 \item KL model (mirror Dubovsky model) $(0,1)$.
\end{itemize}

Let us mention here that necessity of two scaling parameters $\xi$ and $\zeta$ instead of one critical exponent comes from the critical cases when one of the two parameters vanish. Also, the need of two parameters can be made manifest if one considers non-local observables, e.g. Wilson loops \cite{Koroteev:2009qr}.

Let us now consider scalar field in the bulk
\[\label{eq:ScalarFieldAction}
 S = \int\,d^4 x \sqrt{g}\dpod{A}\phi \dpou{A}\phi\,.
\]
We rewrite the metric tensor \eqref{eq:generalansatz} in a different form which is more convenient for the calculations in this section
\[
 \frac{ds^2}{L^2} = -u^{-2\xi}dt^2+u^{-2\zeta}d\textbf{x}^2 + \frac{du^2}{u^2}\,.
\]
Euler-Lagrange equations for \eqref{eq:ScalarFieldAction} read
\[\label{eq:ModeEquation}
 \phi''-\frac{a-2b-1}{u}\phi'+(E^2u^{2(\xi-1)}-\textbf{p}^2u^{2(\zeta-1)})\phi=0\,.
\]
Near the boundary $u=0$ a solution takes the following form
\[
 \phi(t,\textbf{x},u)=u^{\Delta_+}\phi_+(t,\textbf{x})+u^{\Delta_-}\phi_-(t,\textbf{x})\,,
\]
where $\Delta_\pm$ are solutions of
\[
 \Delta(\Delta-\xi-2\zeta)=0\,.
\]
However, there is a special configuration of scaling parameters for which the above constraint needs to be modified. Indeed, if $\xi=0,\zeta=1$ or $\xi=1,\zeta=0$ the above equation will have the following form
\[\label{eq:ScalingDimSpecial}
 \Delta(\Delta-2)=E^2\,,\quad \text{or} \quad \Delta(\Delta-1)=-p^2\,.
\]
Solutions of these equations are
\[\label{eq:ScalingDimKL}
 \Delta_\pm = 1\pm\sqrt{1-E^2}\,,
\]
for the KL model and
\[\label{eq:ScalingDimD}
 \Delta_\pm = \half\pm\half\sqrt{1+4p^2}\,,
\]
for the Dubovsky model. We see here that scaling dimension becomes energy(momenta)-dependent. One can now observe that in the above two critical cases scaling dimensions have similar form to those in Lifshitz theory \cite{Kachru:2008yh} but in the massive case. It appears that an energy scale gets generated when we go from noncritical cases to critical cases we are discussing in this section. In what follows we shall derive boundary correlators for both KL and Dubovsky models and will see that they scale accordingly with \eqref{eq:ScalingDimKL, eq:ScalingDimD}.

\paragraph{The KL model.}

The Green function is the solution of \eqref{eq:ModeEquation} with the following boundary condition
\[\label{eq:BCGreen}
 G(E,\textbf{p},\epsilon) = 1\,,
\]
where $\epsilon$ is the cutoff. There is no necessity in the cutoff in noncritical cases when the mass of scalar field is zero. However, in the cases we discuss there is a nontrivial renormalization albeit the field is massless.\footnote{Appearance of this scale was mentioned in \cite{Koroteev:2009xd} where the spectrum of field fluctuations on the brane was investigated.} The normalizable solution of \eqref{eq:ModeEquation} with boundary condition \eqref{eq:BCGreen} reads
\[
 G(E,\textbf{p},u)=\frac{u}{\epsilon}\frac{\text{K}_\nu(|\textbf{p}|u)}{\text{K}_\nu(|\textbf{p}|\epsilon)}\,,
\]
where $\nu=\sqrt{1-E^2}$. Near the boundary $u=0$ the Green function can be expanded as follows
\[
 G(E,p,u)=\left(\frac{u}{\epsilon}\right)^{1-\nu}\left(1+\left(\frac{|\textbf{p}|u}{2}\right)^{2\nu}\frac{\Gamma(-\nu)}{\Gamma(\nu)}+\dots\right)\,.
\]
The boundary correlator in the momentum space is given by \cite{Kachru:2008yh}
\[
 \langle\mathcal{O}(E,\textbf{p})\mathcal{O}(-E,-\textbf{p})\rangle = G(-E,-\textbf{p},u)\sqrt{g}g^{uu}\dpod{u}G(E,\textbf{p},u)\big\vert_\epsilon^{+\infty}\,.
\]
Due to finiteness of the Green function at infinity it acquires its only contribution at $u=\epsilon$. Ignoring divergent terms proportional to $\epsilon^{-1}$ and keeping the leading term in $\epsilon$ we obtain\footnote{See \cite{Kachru:2008yh} where some arguments about discarding divergent terms are presented.}
\[\label{eq:CorrMom}
 \langle\mathcal{O}(E,\textbf{p})\mathcal{O}(-E,-\textbf{p})\rangle =2^{1-2\nu}\epsilon^{2\nu-2}\frac{\Gamma(-\nu)}{\Gamma(\nu)}|\textbf{p}|^{2\nu}\,.
\]
Then, taking into account \eqref{eq:ScalingDimKL} and performing three-dimensional Fourier transformation of the above correlator one has\footnote{We have absorbed $\epsilon^{2\nu-2}$ into the operators by field renormalization}
\[\label{eq:ScalingKL}
 \frac{1}{(2\pi)^{3/2}}\int \mathrm{e}^{i\textbf{p}\textbf{x}}\langle\mathcal{O}(E,\textbf{p})\mathcal{O}(0,\textbf{0})\rangle d^3 p\sim \frac{1}{|\textbf{x}|^{2\Delta}}\,,
\]
which properly reproduces the scaling behavior of the local operator $\mathcal{O}$.

\paragraph{The Dubovsky model.}

Dubovsky model has the following Green function with the same boundary condition \eqref{eq:BCGreen}
\[
 G(E,p,u)=\sqrt{\frac{u}{\epsilon}}\frac{\text{H}^{(1)}_\nu(Eu)}{\text{H}^{(1)}_\nu(E\epsilon)}\,,
\]
where $\nu=\half\sqrt{1+4p^2}$ and $\text{H}^{(1)}_\nu(z)$ is the Hankel function of the first kind. It has the following expansion near the boundary
\[
 G(E,p,u)=\frac{u^{\half-\nu}}{\epsilon^{\half-\nu}}\left(1+i\left(\frac{Eu}{2}\right)^{2\nu}\frac{\pi(1+i\cot(\pi\nu))}{\Gamma(\nu)\Gamma(\nu+1)}+\dots\right)\,.
\]
The two point correlator after absorbing the cutoff has the following form
\[\label{eq:CorrMomT}
 \langle\mathcal{O}(E,\textbf{p})\mathcal{O}(-E,-\textbf{p})\rangle \approx -i 2^{-2\nu}\frac{\pi(1+i\cot(\pi\nu))}{\Gamma(\nu)\Gamma(\nu+1)}E^{2\nu}\sim E^{2\Delta-1}\,.
\]
In the position space after one-dimensional Fourier transformation in the time direction one has accordingly
\[\label{eq:ScalingD}
 \frac{1}{\sqrt{2\pi}}\int \mathrm{e}^{-i E t}\langle\mathcal{O}(E,\textbf{p})\mathcal{O}(0,\textbf{0})\rangle dE \sim\frac{1}{t^{2\Delta}}\,,
\]
that is the scaling behavior is reproduced correctly, however, coordinate dependence is more complicated due to the nontrivial prefactor in \eqref{eq:CorrMomT}.

Note that both in \eqref{eq:ScalingKL} and \eqref{eq:ScalingD} scaling behavior of boundary correlators derived from the bulk action matches scaling dimensions of scalar operators only in the sence described above, i.e. in the spatial directions for the KL model and in the time direction for the Dubovsky model. This happens due to distinguished asymptotic behavior of the bulk solution in these cases \eqref{eq:ScalingDimSpecial}.

\section{Discussion}\label{sec:Disc}

We have shown that the models proposed in \cite{Koroteev:2007yp,Koroteev:2009xd} and \cite{Kachru:2008yh} have the same solution of Einstein equations by providing equivalence of matter fields in the bulk. Particularly, anisotropy caused by the introduction of a one-form flux in the KLM solution corresponds to different pressures along branelike and extra dimensional directions.

In \cite{Kachru:2008yh} the RG behavior of the theory living on the boundary is discussed for a particular choice of metric tensor corresponding to the Lifshitz model. Using standard holography description one may argue that RG flow from UV to IR is related to the motion of a brane from $z=+\infty$ to $z=-\infty$ (in the notation of \eqref{eq:generalansatz}). Then one can recast small perturbations of the boundary theory in terms of small perturbations of the metric. The numerical analysis performed in \cite{Kachru:2008yh} shows that in IR the theory flows into conformal regime and its holographic dual background tends to the anti de-Sitter space. This is a natural and expected result, however, this analysis is yet incomplete since one needs to generalize it for extremal configurations $\xi=0$ or $\zeta=0$ as well as studying the behavior near critical points in more detail. One can also think of the following gravitational interpretation of this phenomena. It appears that many
solutions (see e.g. \cite{Koroteev:2007yp}) in backgrounds of type \eqref{eq:generalansatz} have matter distributions which are localized near
the UV boundary. In these solutions the invariant energy density
\[
\sqrt{-g(z)}\rho(z)\to 0 \quad \text{as} \quad z\to +\infty
\]
vanishes as we approach the IR boundary. The same behavior is observed for the other components of the energy-momentum tensor. Thus the r.h.s of the Einstein equations \eqref{eq:Einst} becomes merely $\Lambda \delta^A_B$ in the IR region. This implies that the metric asymptotes to a conformal form and the background becomes an $AdS$ space at large $z$. This makes construction of gauge/gravity duality in the IR regime possible (see \cite{Koroteev:2009qr} where some calculations in the direction of the IR duality have been outlined).

Probably the most intricate and intriguing issue of the problem in question is embedding of the toy-models discussed in this note into some supergravity construction and further into string theory. But has been done so far does not go beyond the ``bottom-top'' approach; here we only extended the dictionary. Thus in \secref{sec:GGCorr} we derived scaling behavior of the boundary correlators from the bulk action for two critical cases -- KL model and Dubovsky model. Due to special behavior of bulk solution near the boundary scaling properties of the correlators are in the correspondence with scaling dimensions of the local operators only for space-like scaling in the KL model and time-like scaling in the Dubovsky model. Therefore, we have extended our knowledge of gauge/gravity correspondence in theories with broken Lorentz invariance to all known solutions in the bulk. Certainly, this is only a ``tree-level'' statement and in order to
go beyond it one needs to know both string theory solution and its dual field theory on the boundary. This problem so far remains to be a challenge.

\section*{Acknowledgements}

The authors want to thank A. Vainshtein and A. Zayakin for fruitful discussions and making useful remarks on the manuscript. We also want to thank J. Kinney for proofreading the text. PK is grateful to J. Maldacena for useful discussion and ICTP in Trieste, Italy for hospitality.

\bibliography{liv}

\begin{thebibliography}{10}
\ifx\href\asklfhas\newcommand{\href}[2]{#2}\fi
\ifx\arxivref\asklfhas\newcommand{\arxivref}[1]{\href{http://arxiv.org/abs/#1}%
{#1}}\fi
\ifx\doiref\asklfhas\newcommand{\doiref}[2]{\href{http://dx.doi.org/#1}{#2}}\fi
\raggedright
\small
\parskip 0pt

\bibitem{Koroteev:2007yp}
P.~Koroteev and M.~Libanov,
\textit{``{On Existence of Self-Tuning Solutions in Static Braneworlds without
  Singularities}''},
\textsf{\doiref{10.1088/1126-6708/2008/02/104}{JHEP~0802,~104~(2008)}},
\texttt{\arxivref{0712.1136}}.
%
\bibitem{Kachru:2008yh}
S.~Kachru, X.~Liu and M.~Mulligan,
\textit{``{Gravity Duals of Lifshitz-like Fixed Points}''},
\textsf{\doiref{10.1103/PhysRevD.78.106005}{Phys.~Rev.~D78,~106005~(2008)}},
\texttt{\arxivref{0808.1725}}.
%
\bibitem{Coleman:1998ti}
S.~R.~Coleman and S.~L.~Glashow,
\textit{``{High-Energy Tests of Lorentz Invariance}''},
\textsf{\doiref{10.1103/PhysRevD.59.116008}{Phys.~Rev.~D59,~116008~(1999)}},
\texttt{\arxivref{hep-ph/9812418}}.
%
\bibitem{Gorbunov:2005dd}
D.~S.~Gorbunov and S.~M.~Sibiryakov,
\textit{``{Ultra-large distance modification of gravity from Lorentz symmetry
  breaking at the Planck scale}''},
\textsf{JHEP~0509,~082~(2005)},
\texttt{\arxivref{hep-th/0506067}}.
%
\bibitem{Dubovsky:2001fj}
S.~L.~Dubovsky,
\textit{``Tunneling into extra dimension and high-energy violation of Lorentz
  invariance''},
\textsf{JHEP~0201,~012~(2002)},
\texttt{\arxivref{hep-th/0103205}}.
%
\bibitem{Libanov:2005yf}
M.~V.~Libanov and V.~A.~Rubakov,
\textit{``{Lorentz-violation and cosmological perturbations: A toy brane-world
  model}''},
\textsf{JCAP~0509,~005~(2005)},
\texttt{\arxivref{astro-ph/0504249}}.
%
\bibitem{Koroteev:2009xd}
P.~Koroteev and M.~Libanov,
\textit{``{Spectra of Field Fluctuations in Braneworld Models with Broken Bulk
  Lorentz Invariance}''},
\textsf{\doiref{10.1103/PhysRevD.79.045023}{Phys.~Rev.~D79,~045023~(2009)}},
\texttt{\arxivref{0901.4347}}.
%
\bibitem{PhysRevB.23.4615}
G.~Grinstein,
\textit{``Anisotropic sine-Gordon model and infinite-order phase transitions in
  three dimensions''},
\textsf{\doiref{10.1103/PhysRevB.23.4615}{Phys.~Rev.~B~23,~4615~(1981)}}.
%
\bibitem{PhysRevLett.35.1678}
R.~M.~Hornreich, M.~Luban and S.~Shtrikman,
\textit{``Critical Behavior at the Onset of $k\rightarrow{}$-Space Instability
  on the $\lambda{}$ Line''},
\textsf{\doiref{10.1103/PhysRevLett.35.1678}{Phys.~Rev.~Lett.~35,~1678~(1975)}%
}.
%
\bibitem{Randall:1999vf}
L.~Randall and R.~Sundrum,
\textit{``An alternative to compactification''},
\textsf{Phys.~Rev.~Lett.~83,~4690~(1999)},
\texttt{\arxivref{hep-th/9906064}}.
%
\bibitem{Koroteev:2009qr}
P.~Koroteev and A.~V.~Zayakin,
\textit{``{Wilson Loops in Gravity Duals of Lifshitz-like Theories}''},
\texttt{\arxivref{0909.2551}}.
%
\end{thebibliography}
\bibliographystyle{nb}

\end{document}